\documentclass[aps,prl,reprint,amsmath]{revtex4-2}
\usepackage{hyperref}
\usepackage{color}
\usepackage[caption=false]{subfig}
\usepackage{graphicx}

\DeclareMathOperator{\li}{L}
\DeclareMathOperator{\Li}{Li}
\newcommand{\eH}{\mathcal{H}}
\newcommand{\up}{\uparrow}
\newcommand{\dow}{\downarrow}

\begin{document}

\title{Superfluid transition of a ferromagnetic Bose gas}

\author{Pye Ton How}
\email{pthow@outlook.com}
\affiliation{Institute of Physics, Academia Sinica, Taipei 115, Taiwan}
\author{Sungkit Yip}
\email{yip@phys.sinica.edu.tw}
\affiliation{Institute of Physics, Academia Sinica, Taipei 115, Taiwan}
\affiliation{Institute of Atomic and Molecular Sciences, Academia Sinica, Taipei 115, Taiwan}

\date{\today}

\begin{abstract}
The strongly ferromagnetic spin-1 Bose-Einstein condensate (BEC) has recently been realized with atomic $^{7}$Li.  It was predicted that a strong ferromagnetic interaction can drive the normal gas into a magnetized phase at a temperature above the superfluid transition, and $^{7}$Li likely satisfies the criterion.  We re-examine this theoretical proposal employing the two-particle-irreducible (2PI) effective potential, and conclude that there exists no stable normal magnetized phase for a dilute ferromagnetic Bose gas.  For $^{7}$Li, we predict that the normal gas undergoes a joint first order transition and jump directly into a state with finite condensate density and magnetization.  We estimate the size of the first order jump, and examine how a partial spin polarization in the initial sample affects the first order transition.  We propose a qualitative phase diagram at fixed temperature for the trapped gas.
\end{abstract}

\maketitle

The realization of Bose-Einstein condensation (BEC) in dilute atomic gases \cite{Anderson1995-qo, Davis1995-dt, Bradley1995-lm} led to an explosion of number of superfluid systems that can be studied experimentally, and many of them display exotic, richer physics beyond the simple spinless bose gas model.  Atomic $^{7}$Li gas is such a system: its low-energy hyperfine states form a triplet with total spin one. (For a review on spinor Bose gas, see \cite{Stamper-Kurn2013-cz}.)  Early experiments\cite{Bradley1995-lm} employ magnetic trapping of a spin-polarized gas, while the later all-optical techniques nonetheless rely on magnetic Feshbach resonance to produce a condensate \cite{Gross2008-pj, Pollack2009-fq}.  New technology allows the trapping of unpolarized $^{7}$Li\cite{Huh2020-cy}, which enjoys an internal $SO(3)$ symmetry of spin rotation.  Notably, $^{7}$Li has a spin-dependent, strongly ferromagnetic interaction\cite{Stamper-Kurn2013-cz, Huh2020-cy, Abraham1997-zk, Julienne2014-iv}.

The zero-temperature mean-field ground state of the spin-1 Bose gas is considered in \cite{Ho1998-ug, Ohmi1998-lb}.  Two BEC phases are predicted depending on the interaction parameters: one has a spin dipole moment (``ferromagnetic'') and the other has a quadruple moment (``polar'').  Most prior works on the finite-temperature phase diagram assume a weak spin dependence in interaction\cite{Zhang2004-wb, Kawaguchi2012-up, Lang2014-oz}.  When the relative strength of the spin-dependent part is sufficiently large, Natu and Mueller\cite{Natu2011-xo} predicted a two-step process toward BEC: a ferromagnetic gas first undergoes a bosonic version of Stoner's transition to develop a spontaneous spin dipole moment, and then condenses at a lower temperature.  $^{7}$Li likely satisfies the criterion.

The magnetization is an obvious choice for order parameter characterizing the intermediate ferromagnetic phase (if exists.)  In a second-quantized description, the Bose field itself remains normal, but certain \emph{bilinear} of the field acquires a non-zero expectation value that spontaneously breaks the spin $SO(3)$ but keeps the gauge $U(1)$ intact.  In this regard, it is also akin to the more exotic pair condensate phase\cite{Nozieres1982-jq, Rice1988-lg, Kagan2002-wx, Natu2011-xo}.  Phases characterized by bilinear order parameters (due to various mechanisms) has been proposed in several superfluid systems\cite{Kuklov2004-ib, Babaev2004-op, Kuklov2008-oz, Bojesen2013-lt}.

On the level of Gizburg-Landau theory, this normal-state magnetism is closely related to the time-reversal\cite{Fischer2016-gt,Zeng2021-bt} and lattice rotational\cite{Sun2019-kt, Cho2020-ne} symmetry breaking above unconventional superconductors, as well as the elusive quartet condensation or charge-4e superconductivity\cite{Ropke1998-dy, Berg2009-pw, Fernandes2019-yv, Fernandes2021-sz, Jian2021-oe, Hecker2023-yt}.  We want to especially highlight the similarity between the ferromagnetic Bose gas and the so-called ``vestigial order'' in the context of unconventional superconductivity \cite{Hecker2018-bb, Fernandes2019-yv, Hecker2023-yt}, where the lattice symmetry plays a role similar to the spin $SO(3)$.  Both proposals share the same mechanism: the orders are mediated purely by the fluctuations of the underlying Bose fields (bosonic matter field or Ginzburg-Landau order parameter, respectively.)

In a previous paper, the present authors concluded that the vestigial order scenario cannot be realized in a weak-coupling superconductor\cite{How2023-jn}: the apparent instability toward a vestigial order in fact signals a joint first order transition directly into the appropriate superconducting phase.  For a pseudo-spin-$\frac{1}{2}$ Bose gas, similar theoretical claim of normal-state magnetism was made\cite{Ashhab2005-sy, Radic2014-mr} and refuted\cite{He2020-rq}.  In this letter, we will show that the same claim for a ferromagnetic spin-one Bose gas is also incorrect: when the relative strength of the spin-dependent interaction is sufficiently large, the gas undergoes a joint first order transition into the BEC phase directly, just like its spin-$\frac{1}{2}$ and superconductor cousins.

We will start by reviewing our theoretical method.  Next, we describe the ferromagnetic instability of a homogeneous normal gas, how it is unphysical, and the actual first order normal-BEC transition.  We discuss the experimental signature and propose a qualitative phase diagram for a trapped gas.  Numerical estimations are given for $^{7}$Li under experimental conditions.


\emph{Theoretical model.}  Let $\psi_{s}$ be the annihilation field operator of a spin-1 boson in $m_z$ spin state $s = {\up, \dow, 0}$.  We adopt the notation found in \cite{Ho1998-ug, Nozieres1982-jq} for the Hamiltonian density:
\begin{equation}
\begin{split}
\eH &= \psi^{*}_s \left( -\frac{\hbar^2}{2m}\nabla^2 \right)\psi_s
+ \frac{C_1}{2} \psi^{*}_s \psi^{*}_{s'} \psi_{s'} \psi_{s} \\
&\quad + \frac{C_2}{2}  (\vec{F})_{ss'} \cdot (\vec{F})_{tt'} \, \psi^{*}_{s} \psi^{*}_{t} \psi_{s'} \psi_{t'}.
\end{split}
\end{equation}
Here $\vec{F}$ is the triplet of spin-1 matrices, and repeated indices are summed over.  The effective interaction is due solely to two-body s-wave scattering, an approximation valid in the dilute limit.  This Hamiltonian enjoys an SO(3) symmetry in the spin space, and the total magnetization of the system is conserved.

In terms of s-wave scattering lengths $a_0$ and $a_2$ in the channels of total spin-0 and 2 respectively, the parameters $C_1$ and $C_2$ are
\begin{equation}
C_1 = \frac{4 \pi \hbar^2}{3m} (a_0 + 2a_2); \quad
C_2 = \frac{4 \pi \hbar^2}{3m} (a_2 - a_0).
\end{equation}
Stability requires $C_1 > 0$ and $C_2 > -C_1$, and we restrict our attention to $C_2 < 0$ that favors a ferromagnetic spin moment.  Numerical calculations\ put $C_2/C_1 = -0.46$\cite{Stamper-Kurn2013-cz, Julienne2014-iv} for $^{7}$Li.

We explore the thermodynamic of a uniform gas.  Passing to the grand canonical ensemble, the gas is coupled to the spin-dependent chemical potential $\mu_\up = \mu + h - q$, $\mu_\dow = \mu - h - q$, and $\mu_0 = \mu$.  This describes an overall chemical potential $\mu$ and linear and quadratic Zeeman energies $h$ and $q$ respectively.  We assume that the time scale of typical experiments forbids the relaxation of total magnetization, and $h$ is merely a corresponding Lagrange multiplier for the conserved magnetization\cite{Stenger1998-xq}: the linear Zeeman effect of a physical magnetic field is absorbed into $h$.  We will assume vanishing residual field and set $q = 0$.  The qualitative physics of the first order transition is unaffected by a small $q \neq 0$, and we will comment on the role of $q$ later.

We employ the two-particle irreducible (2PI) effective potential method, essentially a non-relativistic version of the CJT potential\cite{Cornwall1974-fr} \footnote{see SM for derivation}.  The spin-dependent self energy of the boson is treated as the variational parameter in this formalism.  Before proceeding, we adopt a dimensionless form by choosing $k_B T$ and $\lambda_T = \sqrt{2\pi \hbar^2/ m k_B T }$ as the units of energy and length, respectively.  All subsequent numerical values are reported in this unit system.  The dimensionless coupling constants are defined as $c_{1,2} \equiv C_{1,2}/(k_B T \lambda_T^{\,3})$.   Based on the reported parameters\cite{Huh2020-cy}, we adopt $c_1 \approx 0.0024$ and $c_2 \approx -0.0011$ for $^{7}$Li \footnote{See SM for estimation}.  The interaction parameters being small is a direct consequence of the diluteness condition.

The 2PI potential is truncated at two-loop order, and for the normal gas the treatment is identical to a self-consistent Hartree-Fock (HF) approximation\cite{He2020-rq, Van_Schaeybroeck2013-nv}.  The HF self energy is diagonal in the spin basis and momentum-independent, leading to the ansatz that the density of spin-$s$ atom is $\lambda_T^{-3} \Li_{\frac{3}{2}}(e^{-m_s})$, where the dimensionless energy gap $m_s$ becomes the variational parameter.  We introduce the shorthand $\li_s \equiv \Li_{\frac{3}{2}}(e^{-m_s})$.  The dimensionless 2PI potential for normal gas is
\begin{equation}
\begin{split}
\Omega_n &= \sum_s \left[-\Li_{\frac{5}{2}}(e^{-m_s}) - (m_s + \mu_s)\li_s\right] \\
&\quad + (c_1 + c_2)\left(\li_{\up}^2 + \li_{\dow}^2 + \li_{\up}\li_0 + \li_{\dow} \li_0 \right) \\
&\quad + (c_1 - c_2) \li_{\up} \li_{\dow} + c_1 \li_0^2.
\end{split}
\label{normalFreeEnergy}
\end{equation}
(Subscript $n$ stands for normal.)  Once minimized with respect to $m_s$, $\min \Omega_n$ is the negative of pressure in unit of $k_B T/\lambda_T^3$.


\emph{Ferromagnetic instability.}  We obtain the saddle point equations by taking derivatives of $\Omega_n$.  The three equations read
\begin{subequations}
\label{normalSaddlePt}
\begin{align}
\label{normal1}
m_\up + \mu_\up - (c_1 + c_2)(2\li_\up + \li_0) - (c_1-c_2) \li_\dow & = 0; \\
\label{normal2}
m_0 + \mu_0 - (c_1 + c_2) (\li_\up + \li_\dow) - 2 c_1 \li_0 &= 0; \\
\label{normal3}
m_\dow + \mu_\dow - (c_1 + c_2)(2\li_\dow + \li_0) - (c_1-c_2) \li_\up & = 0.
\end{align}
\end{subequations}

We consider first $h \rightarrow 0^{-}$.  There is a symmetric branch of solution with $m_s = m(\mu)$ for all $s$, implicitly given by
\begin{equation}
m + \mu - (4c_1 + 2c_2)\, \Li_{\frac{3}{2}}(e^{-m}) = 0.
\end{equation}
The solution $m(\mu)$ is positive and monotonically decreasing, ending at the mean-field critical point $\mu_c = (4 c_1 + 2c_2) \, \zeta(3/2)$ where $m(\mu_c) = 0$.  For $^{7}$Li under experimental conditions $\mu_c \approx 0.012$.

Take the difference of \eqref{normal1} and \eqref{normal3}:
\begin{equation}
m_\up - m_\dow = (c_1 + 3c_2) \, (\li_\dow - \li_\up).
\end{equation}
It becomes possible to have $m_\up \neq m_\dow$ if $c_2/c_1 < -1/3$, a regime we dub \emph{deep ferromagnetic}.  (The ratio is  $-0.46$ for $^{7}$Li.)  Equivalently the criterion is $2a_0 > 5a_2$.  Linearizing this equation, one obtains the implicit condition for ferromagnetic instability at $\mu = \mu_{\text{ins}}$ along the symmetric branch:
\begin{equation}
\Li_{\frac{1}{2}}(e^{-m(\mu_{\text{ins}})}) = -\left(\frac{1}{c_1 + 3c_2}\right).
\label{insta}
\end{equation}
The RPA spin susceptibility diverges when \eqref{insta} is satisfied.  By analyzing the Hessian matrix, it can be shown that the symmetric branch is a local minimum of $\Omega_n$ when $\mu < \mu_{ins}$, but no longer so for larger $\mu$.  It is tempting to (incorrectly!) identify this instability as an SO(3)-breaking transition into a ferromagnetic phase, but we will presently show that no stable normal solution exists for a dilute gas when $\mu > \mu_{\text{ins}}$\footnote{See SM for possible vestigial phase}.  There is no other competing instability within the range $-c_1 < c_2 < 0$\cite{Natu2011-xo}.

The saddle point equations \eqref{normalSaddlePt} can be solved numerically as-is.  Given the small interaction parameters, however, the so-called critical approximation $\Li_{\frac{3}{2}}(e^{-m_s}) \approx \zeta(3/2) - 2\sqrt{\pi m_s} + O(m_s)$ is appropriate (for this section only).  Within this approximation, the $\sqrt{m_s}$ correction term becomes important when $\mu - \mu_c$ and $m_s$ are of order $O(c_1^2)$, and this non-analytic term drives the instability.

\begin{figure}
\includegraphics[width = 0.45\textwidth]{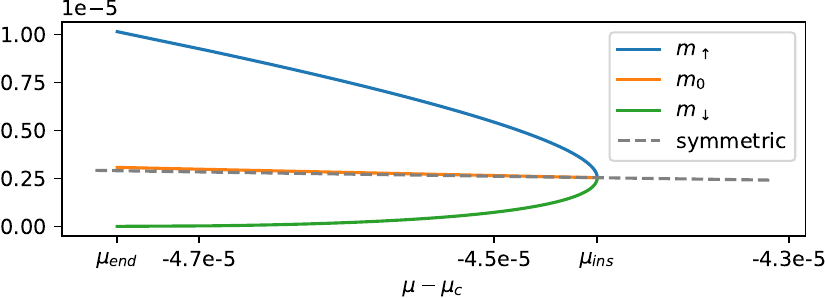}
\caption{\label{NormalSolutions} The symmetric and vestigial ferromagnetic solutions at $h=0$ are plotted for $^{7}$Li.  The vestigial solution ends when $m_\dow = 0$.}
\end{figure}

With the critical approximation, \eqref{normalSaddlePt} admits a closed-form solution when $h \rightarrow 0^{+}$.  The instability occurs at $\mu_{\text{ins}} - \mu_c \approx \pi(c_1 + 3c_2)(7c_1 + c_2) = -4.4 \times 10^{-5}$ for $^{7}$Li.  See Fig \ref{NormalSolutions}.  There is a vestigial ferromagnetic branch connected to the instability, but it lies on the wrong ($\mu < \mu_{\text{ins}}$) side, and is not a minimum of $\Omega$.  The branch ends when $m_\dow = 0$ at some $\mu_{\text{end}} < \mu_{\text{ins}}$.  This so-called ferromagnetic end point is not physical by any means, but will be of some importance in the discussion of the BEC phase.  For $\mu > \mu_{\text{ins}}$, there exists no (meta-)stable normal solution at all.  There must be another global minimum of $\Omega$, and the gas must make a first order jump to this (necessarily superfluidic) state at some $\mu < \mu_{\text{ins}}$.

Our grand canonical approach makes this correct picture apparent.  If one imposes a uniform density constrain instead, it seems at first that the gas enters a ferromagnetic phase as the density is raised\cite{Natu2011-xo}; one needs to work harder to see that the vestigial branch has a negative compressibility and is unstable, as pointed out in \cite{He2020-rq} for the spin-half case.

\begin{figure}
\includegraphics[width = 0.48\textwidth]{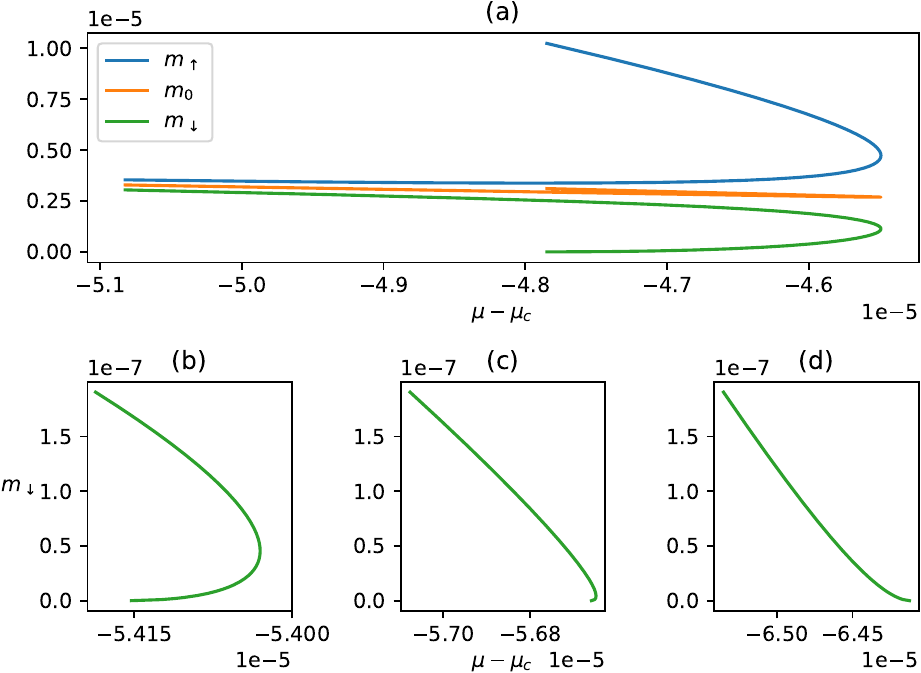}
\caption{\label{ZeemanNormal} We plot the solution with $h < 0$.  (a) Three $m_s$ components for a relatively small $\vert h\vert = 3\times10^{-8}$.  The u-turn is the ferromagnetic instability.  To illustrate the suppressing of instability, we plot $m_\dow$ for (a) $\vert h\vert = 7\times10^{-7} < h_t$, (b) $\vert h\vert = 1\times10^{-6} \lesssim h_t$, and (c) $\vert h\vert = 2\times10^{-6} > h_t$.
}\end{figure}

Without loss of generality, we next consider the case of $h < 0$.  This explicitly breaks the $SO(3)$ symmetry.  See Fig \ref{ZeemanNormal}(a).  For sufficiently small $h$, a unique normal branch evolves from the combination of the $\mu < \mu_{\text{ins}}$ portion of symmetric branch and the vestigial branch.  The solution is still multivalued, and this u-turn is the surviving ferromagnetic instability.  Above a threshold $\vert h \vert > h_t$, the instability is eliminated, and a mean-field-like BEC transition should take place at the gapless ($m_\dow = 0$) end point.  See Fig \ref{ZeemanNormal}(b-d).  For $^{7}$Li gas under experimental condition, we numerically find the threshold $h_t \approx 1.2 \times 10^{-6}$.

Experimentally, this picture of a joint first order transition manifests as coexistence of normal and superfluid phases in a trapped quantum gas.  To estimate the discontinuity across the phase boundary, we need to extend \eqref{normalFreeEnergy} to incorporate the superfluid order.


\emph{Superfluid solution.}  When $\vert h \vert > h_t$, the normal solution sees no instability until $m_\dow$ reaches zero.  From here we expect a continuous transition into the $U(1)$-breaking BEC state as indicated by familiar renormalization group arguments\cite{Zinn-Justin2002-rh}; the gapless end point is also the onset of a BEC branch.  By continuity, when $\vert h \vert < h_{t}$ we also expect the end point of the normal branch to mark the onset of a BEC branch.  The ferromagnetic instability, however, prevents the gas to continuously follow the path.  Before hitting the instability, the gas must makes a first order jump from normal to BEC phase.

\begin{figure}
\includegraphics[width=0.48\textwidth]{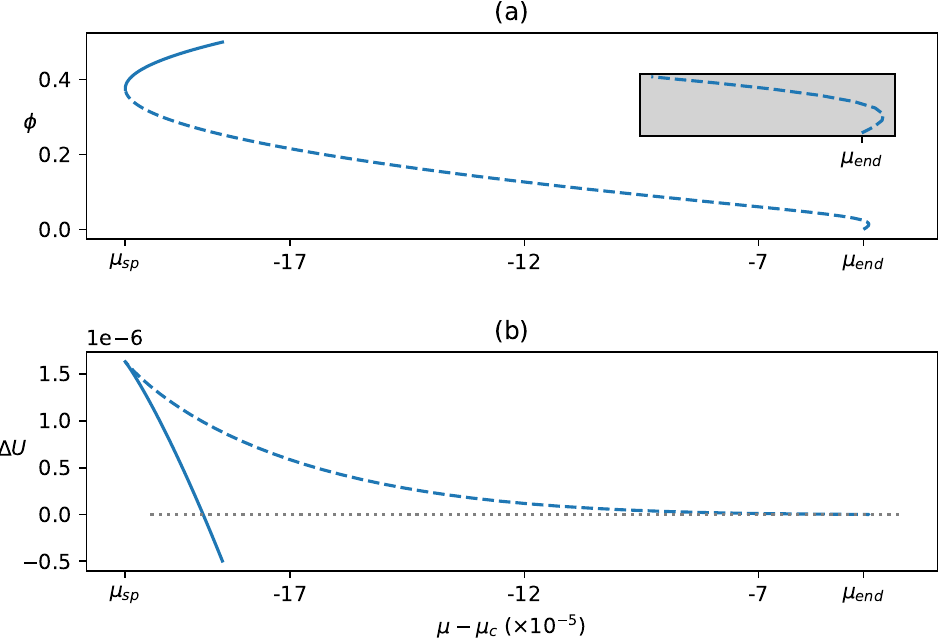}
\caption{\label{BEC}(a) The solution to the saddle point equations with $\phi \neq 0$.  Only the solid part corresponds to local minimum of $\Omega$, and the turning point $\mu_{\text{sp}}$ is the spinodal point.  The insert is the close-up view of the solution near $\mu_{\text{end}}$: one sees the small ``u-turn'' identified as the infrared artifact of the HF approximation.  (b) The plot of $\Delta U \equiv U_b - U_n$.  The first order transition occurs when $\Delta U = 0$.}
\end{figure}

The original work on the CJT potential\cite{Cornwall1974-fr} already provides a general prescription to treat a BEC order.  Concentrating on $h \rightarrow 0^{-}$, we assume the ansatz for the BEC order $ \langle \psi_\up \rangle = \langle \psi_0 \rangle = 0$ and $\langle \psi_\dow \rangle = \phi$, in unit of $\lambda_T^{-3/2}$.  We add to \eqref{normalFreeEnergy} the BEC part:
\begin{equation}
\begin{split}
\Omega_{b} &= -\mu \phi^2 + \frac{1}{2}(c_1 + c_2) \phi^4 \\
&\quad + \phi^2 \left[ (c_1+c_2) (2 L_\dow + L_0) + (c_1 - c_2) L_\up \right].
\end{split}
\end{equation}
The total 2PI potential $\Omega = \Omega_n + \Omega_b$ is required to be stationary with respect to $m_s$ and $\phi$: to the right hand side of \eqref{normal1}-\eqref{normal3}, one adds $(c_1-c_2)\phi^2$, $(c_1+c_2)\phi^2$ and $2(c_1+c_2)\phi^2$, respectively; these goes together with the fourth equation
\begin{equation}
\begin{split}
0 &= -\mu \phi + (c_1 + c_2)\phi^3 \\
&\qquad + \frac{\phi}{2} \left[ (c_1+c_2) (2 L_\dow + L_0) + (c_1 - c_2) L_\up \right].
\end{split}
\label{phiCondition}
\end{equation}
As $\phi = 0$ always solves \eqref{phiCondition}, the symmetric and ferromagnetic branches remain stationary solutions.  But now another ("BEC") solution branch with $\phi \neq 0$ emerges.  Let $U_b$ ($U_n$) denotes the corresponding stationary value of $\Omega$ at the BEC (symmetric) solution, respectively, and the first order transition occurs when $U_n = U_b$.  For $^{7}$Li, our result is summarized in Fig \ref{BEC}.  The BEC branch is initially unstable and extends toward the wrong ($\mu < \mu_{\text{end}}$) side of the ferromagnetic end point, but the branch turns around and becomes a local minimum after it touches the spinodal point.  (On the normal side, the spinodal point is the ferromagnetic instability.)  The scenario is reminiscent of the spin-half case explored by He, Gao and Yu \cite{He2020-rq}.

The combination of 2PI potential and HF approximation is well-known to incorrectly predict a first order transition\cite{Baym1977-lm, Pankov2002-po, Hugel2016-bx, How2023-jn} even when one truly expects a second order one, due to the strong infrared fluctuation when the system is almost gapless.  It also weakly violates\cite{Baym1977-lm} the Goldstone or Hugenholtz-Pines theorem\cite{Watabe2021-mn}.  The manifestation of this infrared problem is the emergence of the non-analytic $\sqrt{m_s}$ in the small-$m_s$ expansion of $L_s$.  We therefore argue that the spinodal structure found here is physical and not an artifact of the infrared problem, since the first order transition occurs far from the region where the $\sqrt{m_s}$ terms dominate.  Following He, Gao and Yu\cite{He2020-rq}, one would identify the infrared artifact with the small ``u-turn'' in the unstable portion of the BEC branch (see insert of Fig \ref{BEC}(a)).


\emph{Experimental signature.}  Within the local density approximation (LDA), an experiment in a trap can be interpreted as a sampling across a range of $\mu$ at fixed $T$ and $h$.  The first order transition shows up as a spatial discontinuities in total density, density of individual spin, magnetization, condensate density, and superfluid density.  The jump is robust against a finite spin imbalance in the cloud.  To estimate the size of the discontinuity, we calculate (at $h\rightarrow 0^{-}$) the density of each spin component on either side of the first order transition.  On the normal side the density per component is $n_n = 2.592$.  On the BEC side, the (purely spin-down) condensate density is $\phi^2 = 0.237$; the densities $n_s$ of the thermal spin-$s$ cloud are $n_{\up} = 2.531$, $n_0 = 2.573$ and $n_{\dow} = 2.551$, respectively.  These results are only weakly dependent on the physical density in the trap\footnote{See SM for result at a different density}.  One then proceeds to work out the mismatches in various quantities.  For example, total density jumps by $(n_\up+n_0+n_\dow+\phi^2)/3n_n - 1 = 2.3\%$.  Relative magnetization is $(n_\up - n_\dow - \phi^2)/(n_\up+n_0+n_\dow+\phi^2) = 3.7\%$ on the BEC side and zero on the normal side.  Not surprisingly, spin-down density shows the biggest discontinuity: $(n_\dow + \phi^2)/n_n - 1 = 8.3\%$.

In a residual magnetic field $B$, the (dimensionless) quadratic Zeeman shift is estimated to be $q \approx (0.02 \, \text{G}^{-2}) \, B^2$, and it should be compared with the the relevant energy scales at the first order transition.  These are the gaps $m_s$ on either sides, and the single particle condensation energy $\frac{c_1 + c_2}{2\zeta(3/2)}\phi^4$.  The smallest is the symmetric gap on the normal side $m \approx 3.5 \times 10^{-5}$.  We conclude that $B = 220\,\text{mG}$ in Ref \cite{Huh2020-cy} will likely obliterate the first order transition as it shifts the energy of spin-$\pm 1$ particles much too high.  However, shielding of residual field to the order $1\text{mG}$ is routine in experiments, and $q$ is then several orders of magnitude smaller than $m$.  Additionally, $q$ can be controlled independently of $B$, reduced and potentially made negative, via e.g. microwave dressing\cite{Gerbier2006-oa}.  The first order transition in fact survives an arbitrarily negative $q$: in the extreme $q \rightarrow -\infty$ limit, the system becomes effectively two-component and still displays the discontinuous transition\cite{He2020-rq}.

\begin{figure}
\includegraphics[width=0.48\textwidth]{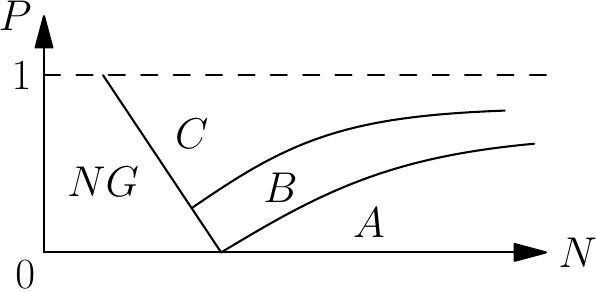}
\caption{\label{in-trap} Qualitative phase diagram of a trapped gas at fixed $T$.  Let $N_s$ be the total number of spin-$s$ atoms, then $N = \sum_s N_s$ and $P = (N_{\up} - N_{\dow})/N$ here.  Phases A, B, C are given in text, and NG stands for normal gas.}
\end{figure}

In a real experiment, the total particle number and magnetization are the external constrains, rather than their conjugates $\mu$ and $h$.  In a trap at fixed temperature, assuming $q=0$, the $h \rightarrow 0^{\pm}$ solution (coexisting ferromagnetic BEC core and unpolarized normal fringe) sets the minimally allowed magnitude of magnetization: a smaller total magnetization can only be accommodated by setting $h=0$ (hence all polarization directions are degenerate,) and allowing the BEC core to have spatially varying polarization.  The normal fringe remains unpolarized.  (The polarization textual of the BEC core is beyond the scope of this work.)  If the magnetization is raised from zero at fixed particle number, the trapped gas exhibits three distinct phases in sequence: discontinuous coexistence of a textured BEC core and an unpolarized normal fringe (phase A), discontinuous coexistence of ferromagnetic BEC and normal fringe (phase B), and continuous coexistence of a ferromagnetic BEC and normal fringe (phase C).  We propose the qualitative in-trap phase diagram Fig \ref{in-trap}.  In phase A, the normal-BEC discontinuities at the coexistence interface are locked at the $h = 0^{\pm}$ values given above.  The Phase B has reduced discontinuities at finite $h$, and phase C has no discontinuity.

As discussed above, to observe the first order coexistence, a positive $q$ must be smaller than the single-particle condensation energy on the BEC side of the transition.  While $q = 0$ is a quantum critical point separating easy-plane and easy-axis ferromagnetisms in a large, homogeneous system, the size of any defect would be extremely large with respect to inter-particle distance given the small $q$.  In a modest-sized cloud, we do not expect polarization textual in phase A to be substantially affected by a small $q \neq 0$ of either sign.  A large and negative $q$ forces the phase A BEC core to separate into domains of up and down spins.


\emph{Conclusion.}  We study the superfluid transition of a dilute ferromagnetic spin-1 Bose gas.  Contrary to a previous claim\cite{Natu2011-xo}, we find that the normal gas cannot support a ferromagnetic phase, regardless of the ratio of interaction parameters.  In the deep ferromagnetic regime where the normal gas does exhibit a ferromagnetic instability upon increasing chemical potential (or density), a ferromagnetic solution exists for the self-consistent HF equation of state, but our grand canonical approach makes it apparent that the solution is thermodynamically unstable.  Instead, a stable BEC solution branch emerges already at a lower chemical potential, and the gas undergoes a joint first order transition into this BEC state \emph{before} hitting the ferromagnetic instability.  In our opinion, such ``vestigial order'' is usually not stabilized in a weakly interacting system, and this is another example.

In a trapped gas, the trap potential translates into spatial variation of the chemical potential $\mu$ within LDA, and the first order transition shows up as the (discontinuous) coexistence of a superfluid core and a normal fringe, with discontinuities in densities and magnetization.  The jump of the majority spin density, the largest of these discontinuities, is estimated to be about $8\%$.  These discontinuities are found to be robust for a range of magnetization in the sample.  We propose the constant-temperature phase diagram Fig \ref{in-trap}.  Too big a residual magnetic field can destroy this physics through the positive quadratic Zeeman shift $q$ in $^{7}$Li, but the qualitative behavior survives if $q$ is tuned negative.

\begin{acknowledgments}
This work is supported by the National Science and Technology Council of Taiwan under grant number MOST 110-2112-M-001-051-MY3.  Additionally, PTH is supported under NSTC 112-2811-M-001-051.
\end{acknowledgments}

\bibliography{BosonMagnetism}

\end{document}


\title{Supplemental materials: Superfluid transition of a ferromagnetic Bose gas}

\author{P. T. How}
\email{pthow@outlook.com}
\author{S. K. Yip}
\email{yip@phys.sinica.edu.tw}

\date{\today}

\maketitle

\section{Estimation of parameters}

We estimate the dimensionless parameters $c_1$ and $c_2$ appearing in the main text for the ultracold $^{7}$Li condensate reported in \cite{Huh2020-cy}.  We remind the reader that the dimensionful parameters are defined as
\begin{equation}
C_1 = \frac{4 \pi \hbar^2}{3m} (a_0 + 2a_2); \quad
C_2 = \frac{4 \pi \hbar^2}{3m} (a_2 - a_0),
\end{equation}
and the dimensionless version goes as
\begin{equation}
c_i = \frac{C_i}{k_B T \lambda_T^{\,3}}, \qquad i = 1, 2.
\label{dimensionless1}
\end{equation}
Here $\lambda_T$ is the thermal wavelength, and $a_{0,2}$ are the s-wave scattering lengths of the angular momentum-zero and two channels, respectively.

We take $a_0 = 23.9 a_B$ and $a_2 = 6.8 a_B$, $a_B$ the Bohr radius\cite{Stamper-Kurn2013-cz,Julienne2014-iv}.  Experimentally, reference \cite{Huh2020-cy} reports peak density $n_{peak} \approx 2.9 \times 10^{19} \operatorname{m}^{-3}$ and average density $n_{avg} \approx 1.5 \times 10^{19} \operatorname{m}^{-3}$.  Within the local density approximation (LDA), the effective local chemical potential decreases as one moves away from the trap center.  In an experiment that probes the superfluid transition, the normal-superfluid coexistent point must be off-center.  So we find it more appropriate to use the average density in our estimation rather than the peak density at the trap center: we set the density at superfluid transition $n_c \approx n_{avg}$.

First of all we test the diluteness condition:
\begin{equation}
a_0 (n_c)^{\frac{1}{3}} \approx 0.0031; \qquad a_2 (n_c)^{\frac{1}{3}} \approx 0.00089.
\label{dimensionless2}
\end{equation}
The gas is indeed very dilute, justifying the effective two-body interaction Hamiltonian in used.  This does not (yet) guarantee that the dimensionless $c_1$ and $c_2$ are small, an issue we will address now.

As a corollary to the dilutenss, the condensation condition for an ideal gas is also a fair approximation:
\begin{equation}
n_c \approx \lambda_T^{\,-3} \, \zeta(3/2).
\label{scale}
\end{equation} 
With this we can rewrite \eqref{dimensionless1}:
\begin{equation}
\begin{split}
c_1 &\approx \frac{2}{3} \left(\frac{n_c}{\zeta(3/2)}\right)^{\frac{1}{3}} (a_0 + 2a_2)  = 0.0024; \\
c_2 &\approx \frac{2}{3} \left(\frac{n_c}{\zeta(3/2)}\right)^{\frac{1}{3}} (a_2 - a_0) = -0.0011. \\
\end{split}
\end{equation}
We are therefore justified in treating $c_1$ and $c_2$ as small parameters.

Incidentally, one also extracts the ratio
\begin{equation}
c_2/c_1 = C_2/C_1 \approx -0.46.
\end{equation}
Thus $^{7}$Li gas sits comfortably within the so-called deep ferromagnetic regime.

The quadratic Zeeman shift due to a magnetic field $B$ is $q = (610\,\text{Hz} \, \text{G}^{-2})\, h\,B^2$ for $^{7}$Li.  Using \eqref{scale} one obtains $T \approx 1.4 \, \mu\text{K}$, which explicitly renders
\begin{equation}
\frac{q}{k_BT} \approx (0.02 \, \text{G}^{-2}) \, B^2.
\end{equation}

\section{2PI Effective Potential}

The 2PI technique has a long history, dating back to the famed Luttinger-Ward potential\cite{Luttinger1960-el}.  The formulation employed in this work is a direct adaptation of the CJT potential\cite{Cornwall1974-fr}.  Conceptually the procedure can be understood as follow.  One first adds the unary and binary source terms to the grand canonical Hamiltonian:
\begin{equation}
\begin{split}
\mathcal{H} &\rightarrow \mathcal{H} 
+ (j_s(x,t) \psi_s(x,t) + \text{c.c}) \\
&\quad + K_{ss'}(x_1, t_1; x_2, t_2) \psi_s^{*}(x_1,t_1) \psi_{s'}(x_2, t_2)
\end{split}
\end{equation}
so that $\langle\psi_s\rangle$ and $\langle\psi_s^{*}\psi_s'\rangle$ become functions of $j_s$ and $K_{ss'}$, and computes the Landau free energy in the presence of these sources.  And then one performs a Legendre transformation with respect to both $j_s$ and $K_{ss'}$ to obtain the effective potential, which is to be minimized with respect to $\langle\psi_s\rangle$ and $\langle\psi_s^{*}\psi_s'\rangle$ at equilibrium.

It turns out that no UV divergences are encountered at the level of approximation, and one may simply set all parameters to their physical, renormalized values from the outset.  Should one want to extend the present work by including divergent Feynman diagrams, the renormalization procedure has also been extensively discussed in the literature\cite{Berges2005-yz,Fejos2008-qt,Carrington2015-la, Rentrop2016-bb}.  However, the crucial different from the relativistic theory is that \emph{all} interactions of the non-relativistic Bose gas are non-renormalizable or irrelevant.  Additional three-body interaction is needed at three-loop level, and $n$-body interaction with arbitrarily high $n$ enters successively at higher loop orders.

We work in imaginary time, and initially in the normal phase with $\langle\psi_s\rangle = 0$.  The full matrix propagator $G_{st}(\tau) = -\langle \operatorname{T} \psi_s(\tau)^{*} \psi_t(0) \rangle$ becomes the variational parameter in the 2PI method.  Given an explicit Zeeman field $h \neq 0$, we expect the full propagator to be diagonal in the $\lbrace\up, 0, \dow\rbrace$ basis, and write down the appropriate ansatz in momentum space:
\begin{equation}
\begin{split}
G^{-1} &= \operatorname{diag}(G_\up^{-1}, G_0^{-1}, G_\dow^{-1}); \\
G_s^{-1} &= i\omega_n - \frac{\hbar^2 k^2}{2m} - m_s
\end{split}
\label{ansatz}
\end{equation}
where $\omega_n$ is the bosonic Matsubara frequency.  The effective gap $m_s$ is just a scalar, refelcting the fact that the Hartree-Fock (HF) correction results in only a uniform energy shift.

\begin{figure}
\includegraphics[scale=2]{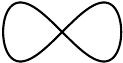}
\caption{\label{2PI1} The (two-loop) Hartree-Fock vacuum diagrams included in the normal part of the 2PI potential $\tilde{\Omega}_n$.  The solid line is the dressed propagator $G_s$.}
\end{figure}

In the normal phase, the only 2PI vacuum diagram up to two-loop level is the family of two-bubble diagrams Fig \ref{2PI1}.  Incorporating these into the calculation, and one writes down the 2PI potential following the CJT formulation:
\begin{widetext}
\begin{equation}
\begin{split}
\tilde{\Omega}_n &= \sum_s \left[ -T \tr \ln(-G_s) - (m_s + \mu_s) \tr (-G_s) \right] \\
&\quad + (C_1 + C_2)\left[\tr(G_\up)^2 + \tr(G_\dow)^2
+ \tr(G_\up)\tr(G_0) + \tr(G_\dow)\tr(G_0) \right] \\
&\quad + (C_1 - C_2) \tr(G_\up)\tr(G_\dow) + C_1 \tr(G_0)^2.
\end{split}
\label{normal2PI}
\end{equation}
\end{widetext}
Here $T$ is the temperature and $\tr$ is short for $T \sum_{\omega_n} \int \frac{d^3k}{(2\pi)^3} e^{-i \omega_n 0^{+}}$: the convergence factor $e^{-i \omega_n 0^{+}}$ regularizes the sum over Matsubara frequencies\cite{Stoof2009-wp}.  The quantity $\tilde{\Omega}_n$ has the dimension of an energy density.  The integrals $\tr \ln (-G_s)$ and $\tr G_s$ yield polylogarithms as are well-known.  Dividing the whole expression by $T \lambda_T^3$ to make it dimensionless, and we recover $\Omega_n$ (3) in the main text.

\begin{figure}
\includegraphics[scale=0.3]{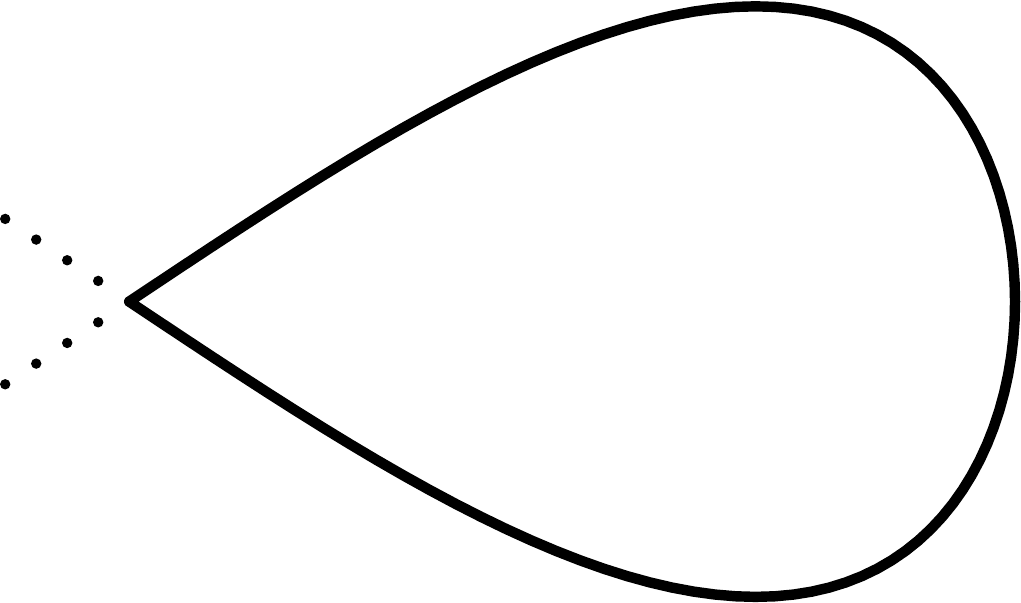}
\caption{\label{2PI2} The additional vacuum diagrams included in the BEC part of the 2PI potential $\tilde{\Omega}_b$.  The dotted line represents the condensate.}
\end{figure}

The general recipe of CJT already allows for a BEC order $\phi_s \equiv \langle \psi_s \rangle \neq 0$.  To begin with, we want to keep the same set of 2PI Hartree-Fock vacuum diagrams Fig \ref{2PI1}.  As these diagrams cannot generate an anomalous part for the self energy, the ansatz \eqref{ansatz} can be kept.  On top of these, the BEC order $\phi_s \neq 0$ generates additional quadratic and cubic interaction vertices for the fluctuating Bose fields.  At the Hartree-Fock level we add the one-loop 2PI diagram Fig \ref{2PI2}.  There is also the classical energy of the field.  The condensate-dependent part reads:
\begin{widetext}
\begin{equation}
\begin{split}
\tilde{\Omega}_b &=
-\mu \left(\phi_s^{*}\phi_s\right)
+ \frac{C_1}{2} \left(\phi_s^{*}\phi_s\right)^2
+ \frac{C_2}{2}  (\vec{F})_{ss'} \cdot (\vec{F})_{tt'} \, \phi^{*}_{s} \phi^{*}_{t} \phi_{s'} \phi_{t'} \\
&\quad + \tr(G_\up) \left[
							2(C_1 + C_2)\vert\phi_\up\vert^2
							+ 2 C_1 \vert\phi_0\vert^2
							+ (C_1-C_2) \vert\phi_\dow\vert^2
					\right] \\
&\quad + \tr(G_0) \left[
							(C_1+C_2)\vert\phi_\up\vert^2
							+ (C_1+C_2)\vert\phi_0\vert^2
							+ (C_1+C_2) \vert\phi_\dow\vert^2
							+ C_1 \left(\phi_\up^{*}\phi_\dow + \phi_\dow^{*}\phi_\up\right)
					\right] \\
&\quad + \tr(G_\dow) \left[
							(C_1-C_2)\vert\phi_\up\vert^2
							+ (C_1+C_2)\vert\phi_0\vert^2
							+ 2(C_1+C_2) \vert\phi_\dow\vert^2
					\right].
\end{split}
\end{equation}
\end{widetext}
Again, one divides $\tilde{\Omega}_b$ by $T \lambda_T^{-3}$ to make it dimensionless.  Note that $\phi_{s}$, being the square root of a number density, should also be rescaled by $\lambda_T^{-3/2}$ to be made dimensionless.  Taking the ferromagnetic ansatz $\phi_{\up} = \phi_0 = 0$, and we recover $\Omega_b$ (8) in the main text.

\section{Stability of the magnetized phase}

Within the critical expansion, which is good for small $c_1$ and $c_2$, we find that the normal, magnetized solution branch is always unstable.  If we take the saddle point equations (4) in the main text at face value and allow $c_1$ and $c_2$ to have arbitrarily big magnitude, the branch can become stable.

We adopt a Ginzburg-Landau approach to analyze the problem, similar to what was done in the authors' previous work on vestigial nematic order above a superconductor\cite{How2023-jn}.  Let $m(\mu)$ be the solution of the symmetric branch (5) in the main text. The deviation from this solution is parameterized by
\begin{equation}
\begin{split}
M &= \frac{1}{3}(m_\up + m_0 + m_\dow) - m(\mu); \\
D &= \frac{1}{2}(m_\up - m_\dow); \\
Q &= \frac{1}{6}(m_\up + m_\dow - m_0).
\end{split}
\end{equation}
The normal phase 2PI potential $\Omega_n$ is then expanded in powers of $M, D, Q$.  Schematically, omitting coefficients, one has
\begin{equation}
\Omega_n \sim \Omega_0 + D^2 + D^2 M + D^2 Q + D^4 + MQ + M^2 + Q^2 + \dots
\end{equation}
One should regard $M, Q$ as $O(D^2)$ for a consistent power counting.  Truncating $\Omega$ as indicated, we first note that vanishing of the coefficient to $D^2$ reproduces exactly the instability condition (6) in the main text.  The truncated saddle point equations are effectively quadratic and can be solved analytically.  The approximation should be accurate in the immediate vicinity of the ferromagnetic instability.  One can then check if and when the magnetized solution corresponds to a (local) minimum of $\Omega_n$ using standard techniques.

Not surprisingly,  the magnetized solution is a local minimum if and only if it exists on the higher $\mu$ (or higher density) side of the instability.  The ratio $k = -(c_1+3c_2)/c_1$ is important; for deep ferromagnetic regime $k > 0$.  When $k \rightarrow 0^+$, we find that a (meta)-stable magnetized phase exists if $c_1 > -\frac{9}{20} \zeta(\frac{1}{2})^{-1} \approx 0.308$.  Somewhat surprisingly, this threshold value of $c_1$ \emph{increases} as $k$ increases.  The threshold diverges when $k = (17 + 3\sqrt{201})/19 \approx 3.13$, and the magnetic phase ceases to exist for $k$ above this value.

The finding here is inline with previous works on vestigial order of unconventional superconductors.  The present authors found that no stable vestigial order exists within a weak-coupling Ginzburg-Landau description\cite{How2023-jn}.  In another study employing a lattice regularization scheme for the loop integrals, the vestigial phase was found to be stabilized only by a very strong interaction\cite{Fischer2016-gt} (and see \cite{How2023-jn} for the interpretation of the strong coupling).

For the Bose gas, we note that the threshold for $c_1$ is very far from the dilute limit.  The two-body effective Hamiltonian is not even applicable, and this stability issue is moot.  Alternatively, the vestigial phase may be stabilized in the large-$N$ limit\cite{Hecker2023-yt}.

\section{Numerical result at a different density}

We report additional numerical results computed with $c_1 = 0.0015$ at the same ratio $c_2/c_1 = -0.46$. 
From \eqref{dimensionless2}, this corresponds to a density that is one quarter of what is adopted everywhere else in this work.

At the first order transition, on the normal side the dimensionless density per component is $n_n = 2.5994$.  On the BEC side, the (purely spin-down) condensate density is $\phi^2 = 0.1471$; the densities $n_s$ of the thermal spin-$s$ cloud are $n_\up = 2.5617$, $n_0=2.5880$ and $n_\dow = 2.5736$, respectively.  The relative magnitudes of the discontinuities decrease: e.g. the spin-down density jumps by $(n_\dow + \phi^2)/n_2 - 1 = 5.1\%$.  We note that the relative changes of these magnitudes are generally less than half even when the density is slashed by three quarters.  The physics of the first order transition is not sensitively dependent on the density.

\bibliography{SM}